\documentclass[prl,twocolumn,showpacs,amsmath,amssymb]{revtex4}
\usepackage{bm}         % Bold Math
\usepackage{graphicx}   % Inclusion of PDF graphics
\usepackage{multirow}   % For table lables
\usepackage{array}      % For table manipulation
\usepackage{ifthen}     % For conditional commands
\newcommand{\mathcommand}[3][0]{\newcommand{#2}[#1]{\ensuremath{#3}}}
\mathcommand[1]{\nth}{#1^{\text{th}}}
\mathcommand{\tcycle}{t_{\text{c}}}
\newcommand{\hamdens}[1][]{\ensuremath{\mathcal{H}_{\text{#1}}}} %symbol for hamiltonian
\renewcommand{\vec}[1]{\mathbf{#1}}
\mathcommand[1]{\aveham}{\mathcal{H}^{(#1)}} % Used for average Hamiltonian terms
\newcommand{\citen}[1]{\onlinecite{#1}} %for compatibility with overcite.sty
\newcommand{\reffig}[1]{Fig.~\ref{#1}}
\mathcommand{\si}{^{29}\text{Si}}
\mathcommand{\p}{^{31}\text{P}}
\newcommand{\sequencename}{CPMG-MREV-16$\times$120}
\newcommand{\latinfont}[1]{\textit{#1}}  % For example, et al.
\newcommand{\bibetal}[2]{{#1 #2 \latinfont{et al.}}}
\newcommand{\ibid}{\latinfont{ibid.}}
\newcommand{\via}{{via}}
\newcommand{\vs}{\latinfont{vs.}}
\newcommand{\mystrut}{\rule{0pt}{2.3ex}} % An x is the height of the LC letter x
\mathcommand[1]{\power}{\times 10^{#1}}
\mathcommand{\tesla}{\text{~T}}
\mathcommand{\cc}{\text{~cm}^3}
\mathcommand{\kelvin}{^\circ\text{K}}
\mathcommand{\percc}{\text{~cm}^{-3}}
\mathcommand{\mol}{\text{~mol}}

\mathcommand{\per}{/\!\!} %Gets rid of that hard space

\mathcommand{\joule}{\text{~J}}
\mathcommand{\eV}{\text{~eV}}
\mathcommand{\meV}{\text{~meV}}
\mathcommand{\watts}{\text{~W}}
\mathcommand{\mHz}{\text{~mHz}}
\mathcommand{\Hz}{\text{~Hz}}
\mathcommand{\kHz}{\text{~kHz}}
\mathcommand{\MHz}{\text{~MHz}}
\mathcommand{\GHz}{\text{~GHz}}
\mathcommand{\THz}{\text{~THz}}
\mathcommand{\PHz}{\text{~PHz}}
\mathcommand{\hr}{\text{~hr}}
\mathcommand{\minutes}{\text{~min}}
\mathcommand{\seconds}{\text{~sec}}
\mathcommand{\ms}{\text{~ms}}
\mathcommand{\us}{\text{~}\mu\text{s}}
\mathcommand{\ns}{\text{~ns}}
\mathcommand{\ps}{\text{~ps}}
\mathcommand{\fs}{\text{~ps}}
\mathcommand{\meter}{\text{~m}}
\mathcommand{\cm}{\text{~cm}}
\mathcommand{\mm}{\text{~mm}}
\mathcommand{\um}{\text{~}\mu\text{m}}
\mathcommand{\nm}{\text{~nm}}
\mathcommand{\angstr}{\text{~\AA}}

\begin{document}
%\preprint{HEP/123-qed}

\title{Coherence Time of a Solid-State Nuclear Qubit}
\author{T. D. Ladd}
\email{tladd@stanford.edu}
\author{D. Maryenko}
\altaffiliation[Currently at ]
            {Max-Planck-Insitut f\"ur Fest\-k\"or\-per\-for\-schung,
            D-70569, Stuttgart, Germany.}
\author{Y. Yamamoto}
\altaffiliation[Also at ]{National Institute of Informatics, Tokyo, Japan.}
\affiliation{Quantum Entanglement Project, ICORP, JST, Edward L.
    Ginzton Laboratory, Stanford University, Stanford, California
    94305-4085, USA}
\author{E. Abe}
\author{K. M. Itoh}
    \affiliation{Department of Applied Physics and Physico-Informatics,
    CREST, JST, Keio University, Yokohama, 223-8522, Japan}%\\
\date{\today}
\begin{abstract}
We report NMR experiments using high-power, RF decoupling
techniques to show that a \si\ nuclear spin qubit in a solid
silicon crystal at room temperature can preserve quantum phase for
$10^9$ precessional periods. The coherence times we report are
longer than for any other observed solid-state qubit by more than
four orders of magnitude.  In high quality crystals, these times
are limited by residual dipolar couplings and can be further
improved by isotopic depletion.  In defect-heavy samples, we
provide evidence for decoherence limited by $1/f$ noise. These
results provide insight toward proposals for solid-state
nuclear-spin-based quantum memories and quantum computers based on
silicon.
\end{abstract}

\maketitle
Quantum information processing devices outperform their classical
counterparts by preserving and exploiting the corre\-lated phases
of their constituent quantum oscillators, which are usually
two-state systems called ``qubits." An increasing number of
theoretical proposals have shown that such devices allow secure
long-distance communication and improved computational
power~\cite{ncbook}. Solid-state implementations of these devices
are favored for reasons of both scalability and integration with
existing hardware, although previous experiments have shown
limited coherence times for solid-state qubits. The development of
quantum error correcting codes
\cite{shor95steane96} and fault tolerant quantum computation
\cite{shor96} showed that large-scale quantum algorithms are still
theoretically possible in the presence of decoherence. However,
the coherence time must be dauntingly long: approximately $10^5$
times the duration of a single quantum gate, and probably longer
depending on the quantum computer architecture~\cite{ncbook}.  The
question of whether a scalable implementation can surpass this
coherence threshold is not only important for the technological
future of quantum computation, but also for fundamental
understanding of the border between microscopic quantum behavior
and macroscopic classical behavior.

\begin{table*}
    \centering
    \caption{Experimental Measurements of Coherence
    Times $T_2$ for Various Qubits.
    These times were all measured by spin-echo Ramsey spectroscopy or
    four-wave-mixing, except where noted.  $Q$ is the product of
    the fundamental qubit frequency $\omega_0/2\pi$ and $\pi T_2$.
    The $\Omega T_2$ column shows the experimental product of the Rabi
    frequency and the coherence time; this provides a more realistic
    measure than $Q$ of the available number of sequential
    single-qubit gates. Technical improvements can increase $\Omega
    T_2$ as far as $Q$ in some architectures, while for others
    $\Omega$ must be limited in order to maintain selective qubit
    control. The $JT_2$ column shows the product of the coherence time
    and the measured or expected qubit-qubit coupling speed; roughly
    the available number of sequential two-qubit gates.
    \label{thetable}}
    \begin{ruledtabular}
    \begin{tabular}[c]{rlrrllll}
      & Qubit & $\omega_0/2\pi$ & $T_2$ & $Q$ & $\Omega T_2$ & $JT_2$ & Ref.\\
      \hline
      \multirow{3}{1ex}{\rotatebox{90}{\makebox[4ex][c]{Atomic~}}} &
        %\mystrut
        \mystrut
        Trapped Optical Ions ($^{40}$Ca$^{+}$) %,$^{199}$Hg$^+$)
        & 412\THz
        & 1\ms
        & $10^{12}$
        & $10^2$
        & $10^1$--$10^3$
        &  \onlinecite{shrgldbreb03skgrdklbehb03}
        \\ %,sl2000}
        & Trapped Microwave Ions ($^{9}$Be$^+$)
        & 1.25\GHz
        & $1\ms$
        & $10^6$
        & $10^3$
        & $10^1$--$10^3$
        & \onlinecite{wmilkm98ldmlbbijlrw03}
        \\ %mmkiw95,,sl2000}
        & Molecular Nuclei in Liquid Solution
        & 500\MHz
        & 2\seconds
        & $10^9$
        & $10^5$
        & $10^2$
        & \onlinecite{vsbysc01}
        \\
    \hline
    \multirow{5}{1ex}{\rotatebox{90}{{Solid-State~}}}
        & \mystrut
        Charge States in Quantum Dots
        & 200--600\THz
        & 40--630\ps
        & $10^5$
        & $10^2$
        & $10^2$
        & \onlinecite{begpks98bhhfkwsf01ptlw03lwsgskdps03}
        \\
        & Josephson-Junction Flux Qubits
        & 6.6\GHz
        & 30\ns
        \footnote{
        \makebox[\textwidth][l]{
        {A recent measurement employing continuous measurement of Rabi
        oscillations indicates a $T_2$ of 2.5\us\ in this system \cite{ioiwgmsmaz03}.}
        }}
        & $10^3$
        & $10^2$
        & $10^1$
        & \onlinecite{moltvl99cnhm03}
        \\
        & Josephson-Junction Charge Qubits  %($\ts{E}{C} > \ts{E}{J}$)
        & 16\GHz
        & 500\ns
        & $10^4$ %30
        & $10^2$
        & $10^4$
        & \onlinecite{mss99vacjpued02pyanat03}%\onlinecite{vacjpued02,pyanat03}
        \\
        & Josephson-Junction Phase Qubits  %($\ts{E}{C} \le \ts{E}{J}$)
        & 16\GHz
        & $5\us$\footnote{
        This result used the decay of low-visibility Rabi
        oscillations.}
        & $10^5$
        & $10^1$
        & $10^4$
        & \onlinecite{yhccw02bxrgsjadlw03}%\onlinecite{yhccw02,bxrgsjadlw03}
        \\
%      Electron Spins Bound to Si:P
%        & 10\GHz
%        & 3\ms
%        & $10^7$
%        & ?
%        & \citen{tlar03,vywjbrmd2000} \\
%        & N-V Color Centers in Diamond
%        & 120\MHz
%        & 0.1\ms
%        & $10^4$
%        & $10^4$
%        & $10^2$
%        & \citen{shlbc02}\\
%      Spin in Quantum Dots &
%       & 10\GHz
%       & 20\ns
%
%        & 300
%        & ?
%        & ? \\
      & \si\ Nuclei in Solid Silicon
      & 60\MHz
      & 25\seconds
      & $10^9$
      & $10^6$
      & $10^4$
      & \onlinecite{lgyyai02},\footnote{As shown in this Letter.}
      \\
    \end{tabular}
    \end{ruledtabular}
\end{table*}

Experimentally observed coherence times ($T_2$) for various qubit
implementations are shown in Table \ref{thetable}.  The atomic
systems shown --- trapped ions and molecular nuclei in liquid
solution --- have already been employed for small quantum
algorithms~\cite{grlbehscb03,vsbysc01}; not coincidentally, they
show very high values of $Q$, the product of the qubit frequency
$\omega_0/2\pi$ and $\pi T_2$. Most solid-state qubits show
smaller values of $Q$; as we demonstrate in this Letter, however,
the long coherence times we observe in solid-state \si\ nuclei
afford them a $Q$ and $\Omega T_2$ as high or higher than atomic
systems, indicating this system's promise for solid-state quantum
computing.

Many promising qubits and coherence time measurements are not
mentioned in Table~\ref{thetable} either because reliable
experimental data are not available, or because the existing
experimental data are taken under conditions not sufficiently
similar to the corresponding quantum computer architecture. For
example, electron spins bound to phosphorous donors in pure,
isotopically depleted silicon also show promisingly long coherence
times~\cite{tlar03}. However, quantum computer architectures
employing semiconductor electron qubits~\cite{ld98,vywjbrmd2000}
require semiconductor alloying and metallic nanostructures for
wave-function engineering.  These additions, rather than bulk
phenomena, are likely to dominate decoherence for these qubits.
Likewise, Kane has proposed an architecture for silicon-based
quantum computing in which embedded \p\ \emph{nuclear} spins are
cooled, controlled, and measured with the engineered wave
functions of donor electrons~\cite{kane98}. Charge fluctuations
couple to the nuclei \via\ the hyperfine interaction, leading to a
nuclear decoherence source not necessarily observed in bulk
experiments.  However, we believe that bulk solid-state nuclei
such as those studied here can be used for quantum computing
without strongly perturbing their local environment.
Reference~\citen{lgyyai02} shows one possible architec\-ture for
quan\-tum com\-put\-ing in nearly-pure, bulk crystalline silicon,
motivating our measurements to explore limits in the coherence
time of bulk \si\ nuclei.

The qubit in this study is a thermal ensemble of $N$ \si\ nuclei,
which are spin-1/2, in a single crystal of silicon. All other
isotopes of silicon are spin-0.  If the \si\ nuclei were uncoupled
and controlled by identical magnetic fields, the system would have
only two accessible degenerate energy levels. However, the dipolar
coupling lifts the degeneracy and takes superpositions of the two
qubit states into much larger, entangled superpositions of all the
$2^{N}$ available states.  These dynamics are nearly independent
of temperature.  The qubit information is not completely lost by
this dipolar scrambling, since information only leaks from the
$2^N$ nuclear states to the lattice on a timescale much slower
than the dipolar dynamics. The quantum information can be
``refocussed" by decoupling pulse sequences such as WAHUHA
\cite{whh68} and its offspring, which periodically reverse the dipolar
evolution, extending $T_2$ by several orders of magnitude.

\begin{figure*}
\includegraphics[width=5.5in]{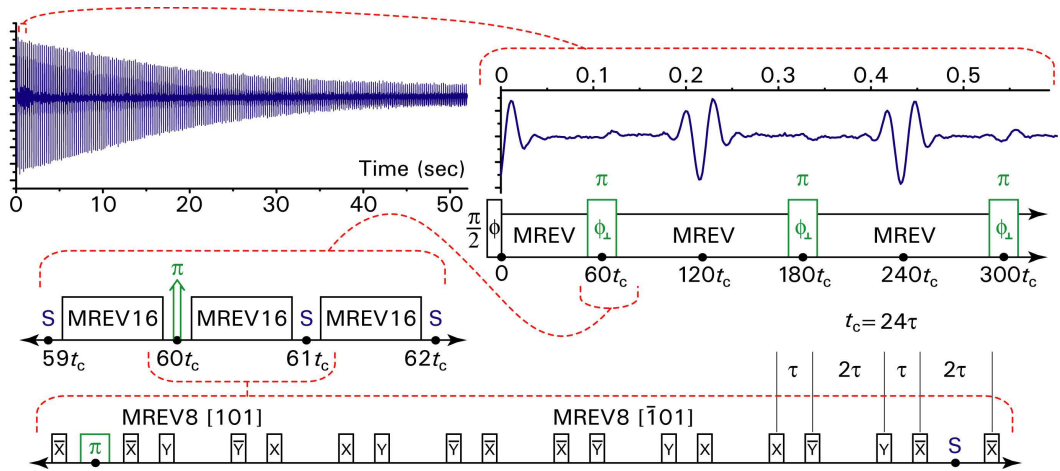}
\caption{Schematic of the \sequencename\ pulse sequence.  The echoes shown
    in the upper left and expanded in the upper right are data from an
    isotopically natural single crystal of silicon.  These are obtained by first exciting
    the sample with a $\pi/2$-pulse of arbitrary phase $\phi$, decoupling with the
    MREV-16 sequence shown in detail on the bottom line, and refocussing with $\pi$-pulses
    of phase $\phi_\perp=\phi+\pi/2$ every 120 cycles of MREV-16.  The magnetization is
    sampled once per MREV-16 cycle in the windows marked with an \textsf{S}.
%graph is from pulses.ai
\label{sequence}}
\end{figure*}

The nuclear dynamics during such sequences are usually described
using average Hamiltonian theory (AHT), in which the effects of
the periodic RF field on the system Hamiltonian are expanded in
powers of $\tcycle$, the sequence period~\cite{mehring83}. The
system Hamiltonian for a nuclear dipolar solid in the rotating
frame may be written as
\begin{equation}
\hamdens=-\sum_j\hbar \omega_j I^z_j - \sum_{jk} D_{jk}
\left[\vec{I}_j\cdot\vec{I}_k-3I^z_jI^z_k\right],
\label{hamiltonian}
\end{equation}
where $\vec{I}_j$ is the spin vector and $\omega_j$ the shift of
the Larmor frequency from the RF carrier for the \nth{j}\ nucleus.
The dipolar coupling coefficients are
$D_{jk}=\hbar^2\gamma^2(1-3\cos^2\theta_{jk})/2r_{jk}^3$, where
$r_{jk}\cos\theta_{jk}=(\vec{r}_j-\vec{r}_k)\cdot\hat{z}$. The
values of $\omega_j$ and $D_{jk}$ are different for each nucleus
due to the inhomogeneous magnetic field and the random
distribution of the
\si\ nuclei in the crystal lattice. The sequence we employ in this
study consists of 16 properly phased and separated $\pi$/2 pulses,
forming MREV-16, a variant of the MREV-8 sequence~\cite{mrev}.
With perfect pulses, the MREV-8 sequence results in the zero-order
average Hamiltonian $\aveham{0}=-(1/3)\sum_j
\hbar\omega_j(I^z_j\pm I^x_j)$, where the sign of the
$x$-component of the effective field depends on the helicity of
the sequence. The MREV-16 sequence, shown in
\reffig{sequence}, concatenates each helicity leading to
$\aveham{0}=-\sum_j (\hbar\omega_j/3) I^z$.  Although MREV-16 has
reduced spectroscopic resolution over MREV-8, maintaining the
effective field in the $z$-direction allows easier nuclear
control. The inhomogeneous offsets described by $\aveham{0}$ cause
static dephasing, which we periodically refocus by applying
$\pi$-pulses every 120 cycles of MREV-16.  We employ the
Carr-Purcell-Meiboom-Gill (CPMG) phase convention to correct for
$\pi$-pulse errors~\cite{mg58}, as shown in \reffig{sequence}.
Such inserted pulses would likewise be employed in a NMR quantum
computer for decoupling and recoupling of multiple dipolar-coupled
qu\-bits~\cite{lcyy2000}.  We note that the CPMG convention
changes the sequence according to the initial phase of the nuclear
qubit, which is incompatible with the memory of unknown or
entangled quantum states.   However, more complex NMR techniques
such as composite pulses \cite{lf81} are known to allow
phase-independent pulse correction. The MREV-16 sequence and the
observed $T_2$ times are otherwise independent of the initial
nuclear phase.

The experiments were performed using a 7\tesla\ superconducting
sole\-noid NMR magnet (Oxford) and a commercial NMR spectrometer
(Tecmag) operating at 59.575\MHz. The probe was homemade,
employing variable vacuum capacitors (Jennings) and high-power
ceramic capacitors (HEC) to accommodate the pulse powers of
approximately 700\watts\ while still allowing flexible tuning for
frequency and impedance matching as well as phase transient
symmetrization. The RF was detuned about 120\Hz\ from the center
of the nuclear resonance frequency. Pulsed spin locking \cite{psl}
was observed when $\pi$-pulses were applied every 5-10 cycles and
in spin-echo experiments without decoupling, but this effect
diminished as the $\pi$-pulse frequency was reduced. Each echo
time-series was taken in a single measurement after thermal
equilibration for half to five times $T_1$.

 As shown in \reffig{sequence}, this combined
``\sequencename" sequence allows the observation of hundreds of
spin-echoes. In a cylindrical sample of single-crystal silicon
with isotopic content 96.9\% \si, the $T_2$ due to
un\-re\-focus\-sed di\-polar-coup\-lings is 450\us\ for the [001]
orientation, as reported previously for this sample~\cite{vmyi03}.
The
\sequencename\ sequence extends the coherence time in this sample
to nearly 2 seconds.  This experiment used a 6\cm-long, 1.5\cm\
coil wound using 2\mm-diameter wire with variable pitch to improve
RF homogeneity.  The $\pi/2$-pulse duration was 9.6\us\ for this
coil.  In isotopically natural silicon, the coherence time is even
longer due to the scarcity of \si\ in the lattice, as
\si\ is naturally only 4.7\% abundant.  We examined a high-quality
single-crystal sample (Marketech) with less than $5\times
10^{13}\percc$ $n$-type impurities. It was cut into a sphere of
diameter 1.5\cm\ and fit tightly in a constant-pitch coil
approximately 6\cm\ long. The $\pi$/2-pulse duration was 15\us\
for this coil. As shown in \reffig{sequence}, the spin-echoes in
this sample last for as long as a minute, showing a $T_2$ of
$25.0\pm 0.2\seconds$.

\begin{figure}
\centering
\includegraphics[width=3in]{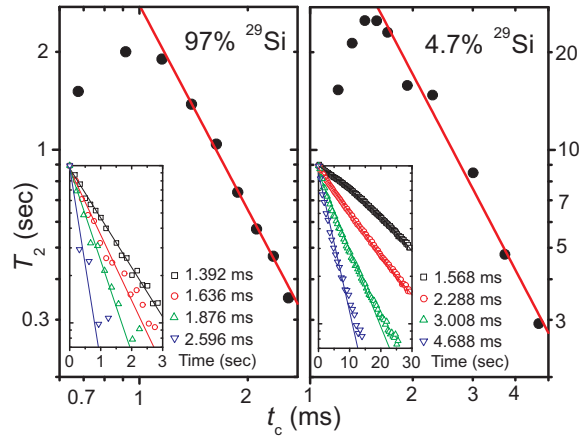}
\caption{
Coherence time \vs\  cycle time.
    The solid line is a fit showing the exponent $-2.09\pm 0.07$ for the
isotopically enhanced sample (left) and $-2.00\pm 0.2$ for the
isotopically natural sample (right).
    The insets show the integrated log-magnitude of the spin-echoes decaying in time
    for a few cycle times.
% Graph is from spheregoodforexport.opj and powerlawforexport.opj
\label{powerlaw}}
\end{figure}

MREV-16 is expected to reduce the effects of the dipolar coupling
to second order in \tcycle\ by AHT, even in the presence of pulse
errors~\cite{mehring83}. Evidence for this order of reduction
comes from the observation that $T_2$ increases quadratically as
$\tcycle$ is decreased (\reffig{powerlaw}).  The spin-echo data
was analyzed by Fourier-transforming each echo and following the
decay of the detuned side-peak, integrated between half-maxima. A
center peak at the RF frequency usually appeared in the spectra
due in part to ringdown effects and in part to nuclear
spin-locking to an effective field, which tips slightly away from
the $z$-axis in the presence of pulse imperfections. The
least-squares fit to a single exponential was always used to
extract the value of $T_2$. The decrease in $T_2$ for small
$\tcycle$ occurs because MREV-16 begins to fail when the ratio of
the pulse-width to the time $\tau$ between pulses approaches
unity~\cite{mehring83}.  The maximum $T_2$ observed (25\seconds)
is not a fundamental result; it may be exceeded using higher-order
pulse sequences or isotopic depletion.

As the strength of the dipolar coupling is further decreased by
isotopic depletion, the dipolar coupling constants $D_{jk}$ become
much smaller than the frequency offsets $\omega_j$. The dominant
second-order dipolar average Hamiltonian term leading to
decoherence is then the dipolar/offset cross term, which scales as
$\aveham{2}\propto
\tcycle^2 |D_{jk}||\omega_j^2|.$ For isotopic percentage $p$ less
than about 10\%, we would expect $T_2^{-1}$ to be proportional to
the dipolar coupling constants, which vary as the inverse cube of
the distance between \si\ isotopes. Correspondingly, we expect
$T_2^{-1}\propto p$, approaching $T_1^{-1}$ as $p\rightarrow 0$,
indicating that the threshold for fault-tolerant quantum computing
could be obtained with reasonably depleted silicon.

\begin{figure}
\centering
\includegraphics[width=2in]{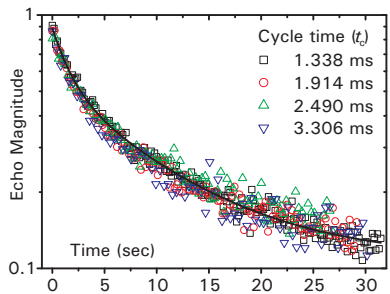}
\caption{Echo decay curves for pure polycrystalline silicon of natural isotopic abundance.
    The solid line is a fit to a double exponential with time constants 1.6 and 9.8 seconds.
    No significant variation in any timescale is observed as \tcycle\ is changed.
% Graph is from file newtauseries.opj
\label{natsi}}
\end{figure}

We did observe another decoherence mechanism that could appear
before the $T_1$ limit in real silicon crystals. In a
polycrystalline sample with shallow impurity content similar to
the single-crystal sample, we found the $T_2$ to be limited to a
constant value of about 8 seconds, independent of $\tcycle$
(\reffig{natsi}). We also examined a $p$-type single crystal
sample with roughly 1-1.5\% \si.  This sample was known to contain
a substantial number of dislocations in addition to $5\times
10^{17}\percc$ aluminum impurities.  Although the data were noisy
due to the reduced signal, $T_2$ was observed to be no longer than
this same timescale of 8 seconds, even after 10-fold averaging.
%Similar
%timescales were also seen for heavily doped commercial silicon
%wafers.

We believe this reduced $T_2$ in polycrystalline silicon is due to
the same low-fre\-quen\-cy noise source that leads to $1/f$ noise
in silicon wafers~\cite{siliconnoise}. This noise is attributed to
charge traps at lattice defects and other deep impurities, which
lead to slow fluctuations of the diamagnetic shielding seen at
nearby nuclei. Free carriers and fixed dipolar paramagnetic
impurities are unlikely to be responsible for this $T_2$, since
these lead to isotropic magnetic noise with correlation times much
shorter than the Larmor period. It follows that $T_2$ due to these
sources would be on the same order of magnitude as $T_1$, but we
measured $T_1$ to be nearly 5 hours, consistent with earlier
studies~\cite{siliconT1}.
%Indeed, the temperature dependence
%of $T_1$ in silicon with less than about $10^{15}\percc$ shallow
%dopants \cite{sl66} suggests an important role for thermally
%activated carriers from deep impurities in nuclear relaxation; the
%low-frequency character of this noise allows $T_2 \ll T_1$, as we
%observe.
The elimination of $1/f$ noise from oxides and interfaces poses a
critical fabrication challenge in quantum computing designs based
on semiconductor impurities~\cite{kane98,vywjbrmd2000} and
Josephson
junctions~\cite{moltvl99cnhm03,mss99vacjpued02pyanat03,yhccw02bxrgsjadlw03},
but this noise is expected to be very small in high-quality bulk
single-crystal silicon at low temperature.
%In fact, previous
%studies~\cite{sl66} have shown that this 5 hour $T_1$ in very pure
%silicon (with less than about $10^{15}\percc$ dopants) is
%independent of doping, magnetic field, and free carrier density,
%and in fact shows a temperature dependence inconsistent with
%either free carriers

Decoupling pulse sequences have been proposed for nuclear memory
in high mobility GaAs/\-AlGaAs heterostructures~\cite{tml03}.  We
caution that the large RF power required to effectively decouple
the ubiquitous nuclear spins in this system may be inconsistent
with millikelvin operation, even if small, high-$Q$ coils and low
power, windowless sequences~\cite{blew} are employed.  For this
reason, we believe isotopically depleted silicon to be a more
promising material for nuclear quantum memory, assuming that
efficient methods for transferring quantum information to and from
its well-isolated nuclei can be found.

This study was sponsored by the DARPA-QuIST program. T.D.L. was
supported by the Fannie and John Hertz Foundation. We acknowledge
N. Khaneja for useful discussions.


\begin{thebibliography}{29}
\expandafter\ifx\csname natexlab\endcsname\relax\def\natexlab#1{#1}\fi
\expandafter\ifx\csname bibnamefont\endcsname\relax
  \def\bibnamefont#1{#1}\fi
\expandafter\ifx\csname bibfnamefont\endcsname\relax
  \def\bibfnamefont#1{#1}\fi
\expandafter\ifx\csname citenamefont\endcsname\relax
  \def\citenamefont#1{#1}\fi
\expandafter\ifx\csname url\endcsname\relax
  \def\url#1{\texttt{#1}}\fi
\expandafter\ifx\csname urlprefix\endcsname\relax\def\urlprefix{URL }\fi
\providecommand{\bibinfo}[2]{#2}
\providecommand{\eprint}[2][]{\url{#2}}

\bibitem[{\citenamefont{Nielsen and Chuang}(2000)}]{ncbook}
\bibinfo{author}{\bibfnamefont{M.~A.} \bibnamefont{Nielsen}} \bibnamefont{and}
  \bibinfo{author}{\bibfnamefont{I.~L.} \bibnamefont{Chuang}},
  \emph{\bibinfo{title}{Quantum Computation and Quantum Information}}
  (\bibinfo{publisher}{Cambridge University Press}, \bibinfo{year}{2000}).

\bibitem[{sho()}]{shor95steane96}
\bibinfo{note}{P. Shor, Phys. Rev. A \textbf{52}, 2493 (1995); A.M. Steane,
  Phys. Rev. Lett. \textbf{77}, 793 (1996).}

\bibitem[{\citenamefont{Shor}(1996)}]{shor96}
\bibinfo{author}{\bibfnamefont{P.}~\bibnamefont{Shor}}, in
  \emph{\bibinfo{booktitle}{Proceedings, {37\textsuperscript{th}} Annual
  Symposium on Fundamentals of Computer Science}} (\bibinfo{publisher}{IEEE
  Press}, \bibinfo{year}{1996}), p.~\bibinfo{pages}{56}.

\bibitem[{shr()}]{shrgldbreb03skgrdklbehb03}
\bibinfo{note}{\bibetal{F.}{Schmidt-Kaler}, Nature \textbf{422}, 408 (2003); J.
  Phys. B: At. Mol. Opt. Phys. \textbf{36}, 623 (2003)}.

\bibitem[{wmi()}]{wmilkm98ldmlbbijlrw03}
\bibinfo{note}{\bibetal{D. J.}{Wineland}, J. Res. Natl. Inst. Stand. Technol.
  \textbf{103}, 259 (1998); \bibetal{D.}{Liebfried}, Nature \textbf{422}, 412
  (2003)}.

\bibitem[{\citenamefont{\bibetal{L. M. K.}{Vandersypen}}(2001)}]{vsbysc01}
\bibinfo{author}{\bibnamefont{\bibetal{L. M. K.}{Vandersypen}}},
  \bibinfo{journal}{Nature} \textbf{\bibinfo{volume}{414}},
  \bibinfo{pages}{883} (\bibinfo{year}{2001}).

\bibitem[{beg()}]{begpks98bhhfkwsf01ptlw03lwsgskdps03}
\bibinfo{note}{\bibetal{N. H.}{Bonadeo}, Science \textbf{282}, 1473 (1998);
  \bibetal{M.}{Bayer}, \ibid\ \textbf{291}, 451 (2001); \bibetal{X.}{Li},
  \ibid\ \textbf{301}, 809 (2003); P. Palinginis, S. Tavenner, M. Lonergan, and
  H. Wang, Phys. Rev. B \textbf{67}, 201307 (2003).}

\bibitem[{\citenamefont{\bibetal{E.}{Il'ichev}}(2003)}]{ioiwgmsmaz03}
\bibinfo{author}{\bibnamefont{\bibetal{E.}{Il'ichev}}}, \bibinfo{journal}{Phys.
  Rev. Lett.} \textbf{\bibinfo{volume}{91}}, \bibinfo{pages}{097906}
  (\bibinfo{year}{2003}).

\bibitem[{mol()}]{moltvl99cnhm03}
\bibinfo{note}{\bibetal{J. E.}{Mooij}, Science \textbf{285}, 1036 (1999); I.
  Chiorescu, Y. Nakamura, C.J.P.M. Harmans, and J.E. Mooij, \ibid\
  \textbf{299}, 1869 (2003)}.

\bibitem[{mss()}]{mss99vacjpued02pyanat03}
\bibinfo{note}{{Y. Makhlin, G. {Sch\"on}, and A. Shnirman, Nature \textbf{398},
  305 (1999); \bibetal{Yu. A.}{Pashkin}, \ibid\ \textbf{421}, 823 (2003);
  \bibetal{D.}{Vion}, Science \textbf{296}, 886 (2002)}}.

\bibitem[{yhc()}]{yhccw02bxrgsjadlw03}
\bibinfo{note}{Y. Yu, S. Han, X. Chu, S-I. Chu, and Z. Wang, Science
  \textbf{296}, 889 (2002); \bibetal{A. J.}{Berkley}, \ibid\ \textbf{300}, 1548
  (2003).}

\bibitem[{\citenamefont{\bibetal{T.D.}{Ladd}}(2002)}]{lgyyai02}
\bibinfo{author}{\bibnamefont{\bibetal{T.D.}{Ladd}}}, \bibinfo{journal}{Phys.
  Rev. Lett.} \textbf{\bibinfo{volume}{89}}, \bibinfo{pages}{17901}
  (\bibinfo{year}{2002}).

\bibitem[{\citenamefont{\bibetal{S.}{Gulde}}(2003)}]{grlbehscb03}
\bibinfo{author}{\bibnamefont{\bibetal{S.}{Gulde}}}, \bibinfo{journal}{Nature}
  \textbf{\bibinfo{volume}{421}}, \bibinfo{pages}{48} (\bibinfo{year}{2003}).

\bibitem[{\citenamefont{Tyryshkin et~al.}(2003)\citenamefont{Tyryshkin, Lyon,
  Astashkin, and Raitsimring}}]{tlar03}
\bibinfo{author}{\bibfnamefont{A.}~\bibnamefont{Tyryshkin}},
  \bibinfo{author}{\bibfnamefont{S.~A.} \bibnamefont{Lyon}},
  \bibinfo{author}{\bibfnamefont{A.~V.} \bibnamefont{Astashkin}},
  \bibnamefont{and} \bibinfo{author}{\bibfnamefont{A.~M.}
  \bibnamefont{Raitsimring}} (\bibinfo{year}{2003}), \eprint{cond-mat/0303006}.

\bibitem[{\citenamefont{Loss and {DiVincenzo}}(1998)}]{ld98}
\bibinfo{author}{\bibfnamefont{D.}~\bibnamefont{Loss}} \bibnamefont{and}
  \bibinfo{author}{\bibfnamefont{D.}~\bibnamefont{{DiVincenzo}}},
  \bibinfo{journal}{Phys. Rev. A} \textbf{\bibinfo{volume}{57}},
  \bibinfo{pages}{120} (\bibinfo{year}{1998}).

\bibitem[{\citenamefont{\bibetal{R.}{Vrijen}}(2000)}]{vywjbrmd2000}
\bibinfo{author}{\bibnamefont{\bibetal{R.}{Vrijen}}}, \bibinfo{journal}{Phys.
  Rev. A} \textbf{\bibinfo{volume}{62}}, \bibinfo{pages}{012306}
  (\bibinfo{year}{2000}).

\bibitem[{\citenamefont{Kane}(1998)}]{kane98}
\bibinfo{author}{\bibfnamefont{B.}~\bibnamefont{Kane}},
  \bibinfo{journal}{Nature} \textbf{\bibinfo{volume}{393}},
  \bibinfo{pages}{133} (\bibinfo{year}{1998}).

\bibitem[{\citenamefont{Waugh et~al.}(1968)\citenamefont{Waugh, Huber, and
  Haeberlin}}]{whh68}
\bibinfo{author}{\bibfnamefont{J.~S.} \bibnamefont{Waugh}},
  \bibinfo{author}{\bibfnamefont{L.~M.} \bibnamefont{Huber}}, \bibnamefont{and}
  \bibinfo{author}{\bibfnamefont{U.}~\bibnamefont{Haeberlin}},
  \bibinfo{journal}{Phys. Rev. Lett.} \textbf{\bibinfo{volume}{20}},
  \bibinfo{pages}{180} (\bibinfo{year}{1968}).

\bibitem[{\citenamefont{Mehring}(1983)}]{mehring83}
\bibinfo{author}{\bibfnamefont{M.}~\bibnamefont{Mehring}},
  \emph{\bibinfo{title}{Principles of High Resolution {NMR} in Solids}}
  (\bibinfo{publisher}{Springer-Verlag}, \bibinfo{year}{1983}).

\bibitem[{mre()}]{mrev}
\bibinfo{note}{P. Mansfield, M. J. Orchard, D. C. Stalker, and K. H. B.
  Richards, Phys. Rev. B \textbf{7}, 90 (1973); W-K. Rhim, D. D. Elleman, and
  R. W. Vaughan, J. Chem. Physics \textbf{59}, 3740 (1973).}

\bibitem[{\citenamefont{Meiboom and Gill}(1958)}]{mg58}
\bibinfo{author}{\bibfnamefont{S.}~\bibnamefont{Meiboom}} \bibnamefont{and}
  \bibinfo{author}{\bibfnamefont{D.}~\bibnamefont{Gill}},
  \bibinfo{journal}{Rev. Sci. Instrum.} \textbf{\bibinfo{volume}{29}},
  \bibinfo{pages}{6881} (\bibinfo{year}{1958}).

\bibitem[{\citenamefont{Leung et~al.}(2000)\citenamefont{Leung, Chuang,
  Yamaguchi, and Yamamoto}}]{lcyy2000}
\bibinfo{author}{\bibfnamefont{D.~W.} \bibnamefont{Leung}},
  \bibinfo{author}{\bibfnamefont{I.~L.} \bibnamefont{Chuang}},
  \bibinfo{author}{\bibfnamefont{F.}~\bibnamefont{Yamaguchi}},
  \bibnamefont{and} \bibinfo{author}{\bibfnamefont{Y.}~\bibnamefont{Yamamoto}},
  \bibinfo{journal}{Phys. Rev. A} \textbf{\bibinfo{volume}{61}},
  \bibinfo{pages}{042310} (\bibinfo{year}{2000}).

\bibitem[{\citenamefont{Levitt and Freeman}(1981)}]{lf81}
\bibinfo{author}{\bibfnamefont{M.~H.} \bibnamefont{Levitt}} \bibnamefont{and}
  \bibinfo{author}{\bibfnamefont{R.}~\bibnamefont{Freeman}},
  \bibinfo{journal}{J. Magn. Reson.} \textbf{\bibinfo{volume}{43}},
  \bibinfo{pages}{65} (\bibinfo{year}{1981}).

\bibitem[{psl()}]{psl}
\bibinfo{note}{E. D. Ostroff and J. S. Waugh, Phys. Rev. Lett. \textbf{16},
  1097 (1966); D. Suwelack and J. S. Waugh, Phys. Rev. B \textbf{22}, 5110
  (1980).}

\bibitem[{\citenamefont{Verhulst et~al.}(2003)\citenamefont{Verhulst, Maryenko,
  Yamamoto, and Itoh}}]{vmyi03}
\bibinfo{author}{\bibfnamefont{A.}~\bibnamefont{Verhulst}},
  \bibinfo{author}{\bibfnamefont{D.}~\bibnamefont{Maryenko}},
  \bibinfo{author}{\bibfnamefont{Y.}~\bibnamefont{Yamamoto}}, \bibnamefont{and}
  \bibinfo{author}{\bibfnamefont{K.~M.} \bibnamefont{Itoh}},
  \bibinfo{journal}{Phys. Rev. B} \textbf{\bibinfo{volume}{68}},
  \bibinfo{pages}{054105} (\bibinfo{year}{2003}).

\bibitem[{sil({\natexlab{a}})}]{siliconnoise}
\bibinfo{note}{R. D. Black, M. B. Weissman, and P. J. Restle, J. Appl. Phys.
  \textbf{53}, 6280 (1982); Phys. Rev. B \textbf{28}, 1935 (1983).}

\bibitem[{sil({\natexlab{b}})}]{siliconT1}
\bibinfo{note}{R. G. Shulman and B. J. Wyluda, Phys. Rev. \textbf{103}, 1127
  (1956); G. Lampel and I. Solomon, C. R. Acad. Sc. Paris \textbf{258}, 2289
  (1964); B. Sapoval and D. Lepine, J. Phys. Chem. Solids \textbf{27}, 115
  (1966).}

\bibitem[{\citenamefont{Taylor et~al.}(2003)\citenamefont{Taylor, Marcus, and
  Lukin}}]{tml03}
\bibinfo{author}{\bibfnamefont{J.~M.} \bibnamefont{Taylor}},
  \bibinfo{author}{\bibfnamefont{C.~M.} \bibnamefont{Marcus}},
  \bibnamefont{and} \bibinfo{author}{\bibfnamefont{M.~D.} \bibnamefont{Lukin}},
  \bibinfo{journal}{Phys. Rev. Lett.} \textbf{\bibinfo{volume}{90}},
  \bibinfo{pages}{206803} (\bibinfo{year}{2003}).

\bibitem[{\citenamefont{Burum et~al.}(1981)\citenamefont{Burum, Linder, and
  Ernst}}]{blew}
\bibinfo{author}{\bibfnamefont{D.}~\bibnamefont{Burum}},
  \bibinfo{author}{\bibfnamefont{M.}~\bibnamefont{Linder}}, \bibnamefont{and}
  \bibinfo{author}{\bibfnamefont{R.}~\bibnamefont{Ernst}}, \bibinfo{journal}{J.
  Magn. Reson.} \textbf{\bibinfo{volume}{44}}, \bibinfo{pages}{173}
  (\bibinfo{year}{1981}).
\end{thebibliography}
\end{document}